\documentclass[aip,reprint,floatfix]{revtex4-1}
\usepackage{amsfonts,amssymb,amsmath,array}

\usepackage{dcolumn}
\usepackage{bm}
\usepackage[final]{graphicx}
\usepackage{graphicx}
\usepackage{epstopdf}
\graphicspath{{./Figures/}} %Setting the graphicspath

\usepackage[usenames]{color} %Per permettere la colorazione dei caratteri 

\usepackage{soul} %per cancellare le frasi

\newcommand{\beq}{\begin{equation}}
\newcommand{\eeq}{\end{equation}}
\newcommand{\bea}{\begin{eqnarray}}   
\newcommand{\eea}{\end{eqnarray}}

\begin{document}

\title{Self-alignment and anti-self-alignment suppress motility-induced phase separation in active systems}

 \author{Marco Musacchio}
	\affiliation{
		Institut f{\"u}r Theoretische Physik II: Weiche Materie,
		Heinrich-Heine-Universit{\"a}t D{\"u}sseldorf, Universit{\"a}tsstra{\ss}e 1,
		D-40225 D{\"u}sseldorf, 
		Germany}

\author{Alexander P.\ Antonov}
	%\email{alexander.antonov@hhu.de}
	\affiliation{
		Institut f{\"u}r Theoretische Physik II: Weiche Materie,
		Heinrich-Heine-Universit{\"a}t D{\"u}sseldorf, Universit{\"a}tsstra{\ss}e 1,
		D-40225 D{\"u}sseldorf, 
		Germany}

\author{Hartmut L{\"o}wen}
	\affiliation{
		Institut f{\"u}r Theoretische Physik II: Weiche Materie,
		Heinrich-Heine-Universit{\"a}t D{\"u}sseldorf, Universit{\"a}tsstra{\ss}e 1,
		D-40225 D{\"u}sseldorf, 
		Germany}

\author{Lorenzo Caprini}
\email{lorenzo.caprini@uniroma1.it}
\affiliation{Physics Department, Sapienza University of Rome, Piazzale Aldo Moro 5 - Rome. }

\begin{abstract}
 In this article, we investigate the impact of self-alignment and anti-self-alignment on collective phenomena in dense active matter. These mechanisms correspond to effective torques that align or anti-align a particle’s orientation with its velocity, as observed in active granular systems.
       In the context of motility-induced phase separation (MIPS) - a non-equilibrium coexistence between a dense clustered phase and a dilute homogeneous phase - both self- and anti-self-alignment are found to suppress clustering.
       Specifically, increasing self-alignment strength first leads to flocking within the dense cluster, and eventually to the emergence of a homogeneous flocking phase. In contrast, anti-self-alignment induces a freezing phenomenon, progressively reducing particle speed until MIPS is suppressed and a homogeneous phase is recovered. 
       These results are supported by scaling arguments and are amenable to experimental verification in high-density active granular systems exhibiting self- or anti-self-alignment.
\end{abstract}

\maketitle

Active systems~\cite{marchetti2013hydrodynamics, Elgeti2015}, which convert energy from the environment into directed motion \cite{bechinger2016active, o2022time}, are characterized by a plethora of spontaneous collective phenomena.
Several macroscopic and microscopic systems, such as animals~\cite{cavagna2014bird, cavagna2017dynamic}, cells~\cite{alert2020physical} and bacteria~\cite{zhang2010collective}, often align their velocities and show synchronized collective motion. The transition from a disordered to global motion is interpreted as a non-equilibrium phase transition, often referred to as flocking~\cite{vicsek2012collective, mahault2019quantitative, chate2020dry, caprini2023flocking, ihle2011kinetic}. 
Starting from the pioneering work of T. Vicsek~\cite{vicsek1995novel}, minimal microscopic models, such as the Vicsek model~\cite{vicsek2012collective, chate2008modeling} or the inertial spin model~\cite{cavagna2018physics}, have been introduced to explain this phenomenon and reproduce experimental data~\cite{cavagna2022marginal}.
Typically, these dynamics assume the existence of effective alignment interactions among different individuals which tend to align or anti-align the orientations of different particles through effective torques~\cite{levis2019activity, zhao2021phases, kursten2025emergent, mangeat2024emergent}.

Another class of active systems, ranging from bacteria to active colloids, typically displays a non-equilibrium phase coexistence between a low-density and a high-density phase~\cite{buttinoni2013dynamical, palacci2013living, van2019interrupted}. This phenomenon is referred to as motility-induced phase separation~\cite{cates2015motility} (MIPS), resembles the typical scenario of a first-order phase transition~\cite{fily2012athermal, Redner2013Structure, gonnella2015motility, solon2015pressure, speck2016collective, caporusso2020motility, maggi2022critical, omar2023mechanical} enriched by several non-equilibrium effects, ranging from spontaneous velocity alignment in the dense phase~\cite{caprini2020spontaneous, caprini2020hidden, kopp2023spontaneous, yang2023coherent} to kinetic temperature difference~\cite{mandal2019motility, petrelli2020effective, hecht2024motility} and anomalous interfacial tension between dense and dilute phases~\cite{bialke2015negative, levis2017active, hermann2019non, patch2018curvature,omar2020microscopic}.
These phenomena have been explained through simple models of particles that interact via volume exclusion and are characterized by persistent dynamics that models particle motility~\cite{bialke2015active, digregorio2018full, klamser2018thermodynamic, dai2020phase, su2021inertia, caprini2022role, de2022collective, de2022motility, kryuchkov2023inertia, ai2024rotational}.

Recently, an upsurge of interest has emerged in the study of active granular matter \cite{scholz2018rotating, walsh2017noise, deblais2018boundaries, lopez2022chirality, baconnier2022selective, siebers2023exploiting, caprini2024emergent} which offers captivating frontiers in soft robotics applications ranging from swarm robotics to spatial exploration or rescue missions. Active granular particles are macroscopic asymmetric objects that show active motion due to local injection of energy due to internal motors \cite{Agrawal2020, leoni2020surfing, tapia2021trapped} or global energy injection generated by the environment, such as a vertically vibrating shaker \cite{aranson2007swirling, kudrolli2008,  deseigne2010collective, kudrolli2010concentration, Koumakis2016, scholz2018inertial, antonov2024inertial}. These systems behave as active particles because they run with a typical speed and a certain degree of persistence as if their motion is sustained by a self-propulsion force. In addition, they show collective phenomena spanning from clustering \cite{chen2023molecular, caprini2024dynamical} to flocking behavior \cite{kumar2014flocking, soni2020phases, casiulis2024geometric, arbel2024mechanical}, as well as anomalous cooling effects~\cite{antonov2025self}.
However, compared to active colloids or bacteria, active granular particles are often characterized by self-alignment or anti-self-alignment mechanisms~\cite{baconnier2024self}, i.e.\ an effective torque that aligns or anti-aligns the particle orientation with its velocity. Contrary to Vicsek-like models, alignment manifests itself as a single-particle mechanism that is usually generated by an asymmetric distribution of propulsive and dissipative forces in the particle body~\cite{baconnier2024self}. 

In a many-body system of particles with non-aligning interactions, the self-alignment mechanism generates a transition from a disordered to a flocking phase~\cite{szabo2006phase, weber2013long, lam2015self, barton2017active, malinverno2017endocytic, giavazzi2018flocking}, even though this mechanism does not directly couple the orientations of different particles. 
Self-alignment is also responsible for further fascinating effects, ranging from orbital motion when particles are confined in a harmonic potential~\cite{dauchot2019dynamics, canavello2024polar} to a reentrant glass transition in a polydisperse active system~\cite{paoluzzi2024flocking}.
Finally, self-alignment can excite spontaneous collective excitations, named collective actuation~\cite{baconnier2022selective}, in a crystal consisting of granular particles subjected to self-alignment. 

Here, inspired by active granular particles, we investigate the effect of self- and anti-self-alignment mechanisms on the collective phenomena of inertial active systems.
Before clustering is suppressed in favor of a homogeneous flocking phase, the self-alignment mechanism promotes a flocking MIPS, with a polarized flocking cluster immersed in a low-density homogeneous liquid.
By contrast, anti-self-alignment induces a freezing effect: it has the primary effect of slowing down each active particle until its motility is so low that clustering cannot occur.

The paper is structured as follows: After introducing the model in Sec.~\ref{sec:model}, we numerically study collective phenomena in interacting active particles subject to self or anti-self-alignment mechanisms in Sec.~\ref{sec:numerics}.
Finally, we present a discussion in the conclusive section.

\section{Model}
\label{sec:model}

\begin{figure*}[!t]
\centering
\includegraphics[width=1\linewidth,keepaspectratio]{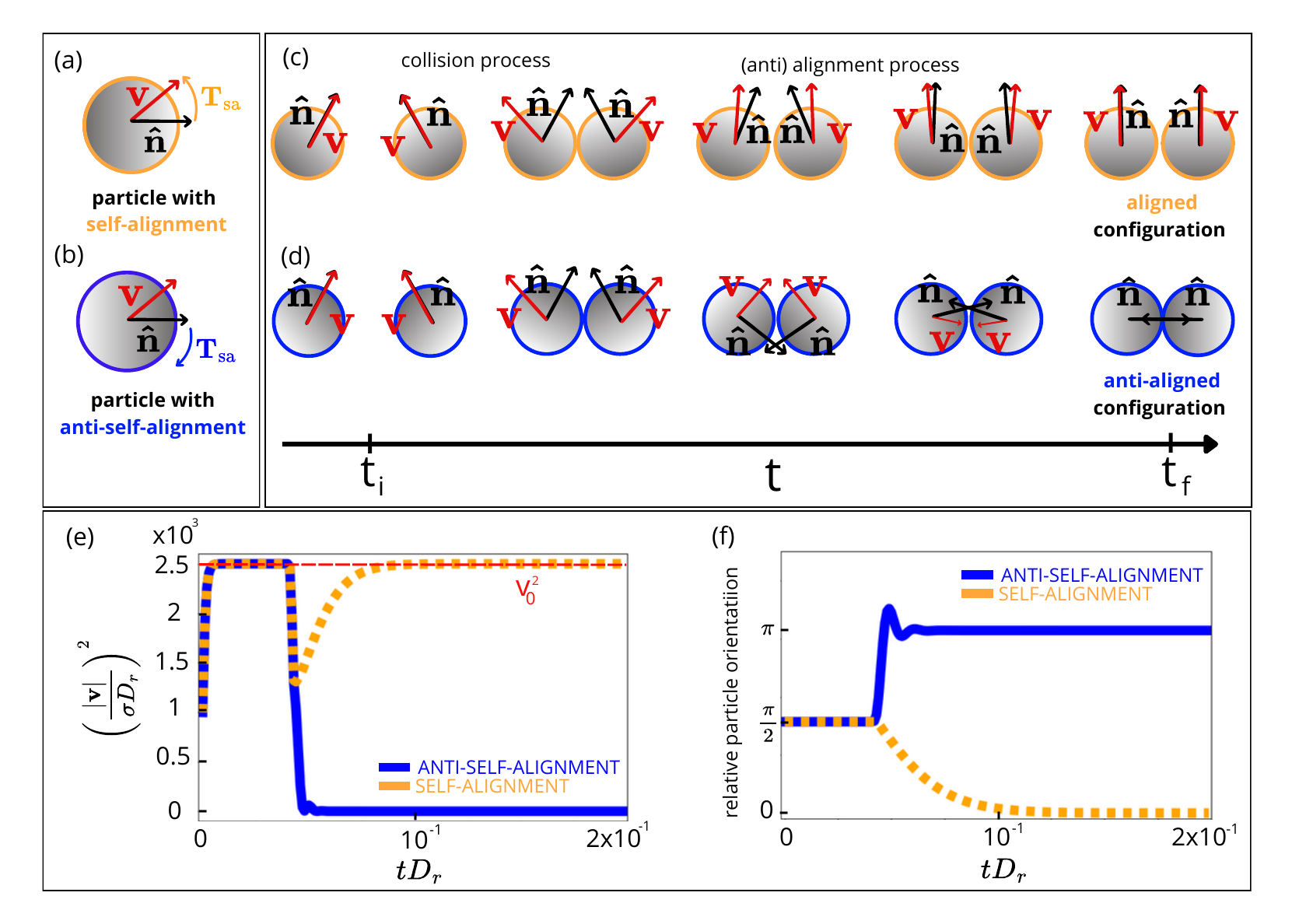}
\caption{
\textbf{Collision mechanism between active particles subject to self-alignment or anti-self-alignment.} (a)-(b) Illustration of active particles governed by self-alignment ($\beta>0$) and anti-self-alignment ($\beta<0$), represented by a circle with the boundary colored orange and blue, respectively. The color gradient in the circles distinguishes between particles governed by self-alignment (a) and anti-self-alignment (b), exerting a torque $\mathbf{T}^{sa}$ on $\hat{\mathbf{n}}$ which tends to align or anti-align it with the velocity $\mathbf{v}$. (c)-(d) Schematic illustration of the collision process as a function of time $t$. Before the impact, two colliding particles have velocities pointing against each other. Immediately after the impact, particle velocities point away because of repulsive interactions, while the orientations remain unchanged. In both self- and anti-self-alignment cases, velocity then starts to restore its initial direction following the propulsion vector $\hat{\mathbf{n}}$. In the self-alignment case (c), the aligning torque tends to align the orientation with the particle velocity. Since ongoing collisions suppress the horizontal component of the velocity, the final configuration results in a perfect vertical alignment between the two particles. In the anti-self-alignment case (d), the orientation rotates away from the velocity direction, thereby progressively slowing down the particles until they are stuck against each other with vanishing velocity. (e) Time evolution of the square velocity $|\mathbf{v}|^2$ during subsequent collisions; (f) time evolution of the relative angle between the orientation of the two particles. Orange and blue lines denote self-alignment and anti-self-alignment, respectively. The dimensionless parameters of the simulations are $M=0.001$, $\text{Pe} = 50$, $D_r J/\gamma_r=10^{-3}$, $B = 10^3$ for the self-alignment and $B = -10^4$ for the anti-self-alignment.}
\label{fig:fig1}
\end{figure*}

We study a two-dimensional system composed of inertial active particles. Beyond the self-propulsion force typical of active particles, the particle orientation is subject to a (anti-) self-alignment mechanism, i.e.\ a torque that tends to align or anti-align the orientation with the particle velocity. The translational dynamics of a particle $i$ with a mass $m$ is described by a stochastic equation for the particle position $\mathbf{r}_i$ and the velocity $\mathbf{v}_i=\dot{\mathbf{r}}_i$, given by
\begin{align}    
    m\dot{\mathbf{v}}_i &= -\gamma \mathbf{v}_i + \gamma v_0\mathbf{\hat{n}}_i + \mathbf{F}_i + \gamma\sqrt{2 D_t}\,\boldsymbol{\xi}_i \,,
\label{eq:eulero_pos}
\end{align}
\noindent where $\boldsymbol{\xi}_i$ is a Gaussian white noise with zero average and unit variance. The terms $\gamma$ and $D_t$ represent the translational friction coefficient and the translational diffusion coefficient due to the environment. 
The particle is active because it is subject to a self-propulsion force $\gamma v_0\hat{\mathbf{n}}_i$ that drives the particle motion along the unit orientational vector $\hat{\mathbf{n}}_i=(\cos\theta_i, \sin\theta_i)$, determined by the orientational angle $\theta_i$. The term $v_0$ represents the constant speed provided by the active force that determines the activity level of the particle. 
Volume exclusion effects are included via repulsive interactions $\mathbf{F}_i$ between the particles which can be expressed as $\mathbf{F}_i = -\nabla_i U_{\rm tot}$. Here, $U_{\rm tot}$ represents a total, pairwise not-aligning potential, $U_{\rm tot} = \sum_{i < j} U(\lvert \mathbf{r}_i - \mathbf{r}_j \rvert)$, where $U(r) = 4\epsilon \left[ \left(\sigma/r\right)^{12} - \left(\sigma/r\right)^{6}\right]$ is the Weeks-Chandler-Andersen (WCA) potential. In this expression, $\epsilon$ denotes the energy scale while $\sigma$ represents the particle diameter. In our simulations, we consider the non-thermal limit $D_t=0$, since in experiments this term is often negligible compared to the active force~\cite{bechinger2016active}.. 

A rigid body in two dimensions is additionally described by an evolution equation for the particle angular velocity, $\omega_i=\dot{\theta}_i$, which is governed by an underdamped equation of motion for the orientational angle $\theta_i$
\begin{align}
%    \dot{\varphi}_i &= \omega_i \\
    \quad J \dot{\omega}_i  &= -\gamma_r \omega_i +  \mathbf{T}^{sa}_i \cdot \hat{\mathbf{e}}_z + \gamma_r \sqrt{2 D_r} \eta_i\,,
\label{eq:eulero_ang}
\end{align}
\noindent where $J$ is the particle moment of inertia, $\gamma_r$ represents the rotational friction coefficient, and $\mathbf{\hat{e}}_z$ is a unit vector orthogonal to the plane of motion. The term $\sqrt{2 D_r} \eta_i$ quantifies the rotational noise acting on the particles, being $\eta_i$ a white noise with zero average and unit variance and $D_r$ the rotational diffusion coefficient. 
This equation of motion includes the (anti-) self-alignment mechanism, through an effective torque $(\mathbf{T}^{sa}_i \times \mathbf{\hat{n}}_i)$ which is modeled as:
\begin{align}
   \mathbf{T}^{sa}_i= \beta (\mathbf{\hat{n}_i} \times \mathbf{v}_i)\,.   
\end{align}
This term aligns the orientation with the velocity when the parameter $\beta$ is positive or anti-aligns with the velocity when $\beta$ is negative. The parameter $\beta$ determines the strength of this (anti-) self-aligning mechanism compared to the random term and introduces the additional self-aligning distance $\gamma_r/|\beta|$. This parameter roughly represents the distance run by a particle after the active force orientation aligns or anti-aligns with the velocity.

The dynamics are governed by three different characteristic times: the persistence time of a single particle's trajectory, $\tau_p = 1/D_r$; the translational inertial time, $\tau_d = m/\gamma$; and the rotational inertial time $\tau_r = J/\gamma_r$.
Rescaling the positions in units of the particle's diameter, $\sigma$, and the time in units of the persistence time, $\tau_p$, the dynamics is governed by several dimensionless parameters (see Appendix~\ref{appendix:A}). The level of activity is quantified by the P\'eclet number, $\text{Pe} = v_0/(D_r \sigma)$, which is given by the ratio between persistence length and particle diameter; the reduced mass $M = D_r m/\gamma$ determines the relevance of inertia compared to the persistence time, while the self-alignment strength $B = \beta \sigma / J D_r$ quantifies the role of the self-alignment mechanism and will be central to our analysis.
Other dimensionless parameters that govern the dynamics are the reduced moment of inertia $D_r J/\gamma_r$, i.e.,\ the ratio between the rotational inertial time and the persistence time and the potential strength $\sqrt{\epsilon/m}/(D_r \sigma)$. 

Simulations are performed considering $N$ particles in a box of size $L$ with periodic boundary conditions, at a fixed packing fraction $\Phi=N\sigma^2 \pi/(4 L^2)=0.5$. We set the P\'eclet number $\text{Pe}=50$ so that motility-induced phase separation is numerically obtained at vanishing self-alignment and small inertia. Here, we vary the self-alignment strength $B$ and the reduced inertial mass $M$ by fixing the remaining dimensionless parameters.
Specifically, the reduced moment of inertia is sufficiently small so that the angular velocity relaxes fast and does not play any role, while the potential strength is kept unitary because this value does not significantly affect the phase diagram as shown in previous studies~\cite{caprini2022role}.

\section{Results}
\label{sec:numerics}

\begin{figure*}[!t]
\centering
\includegraphics[width=1\linewidth,keepaspectratio]{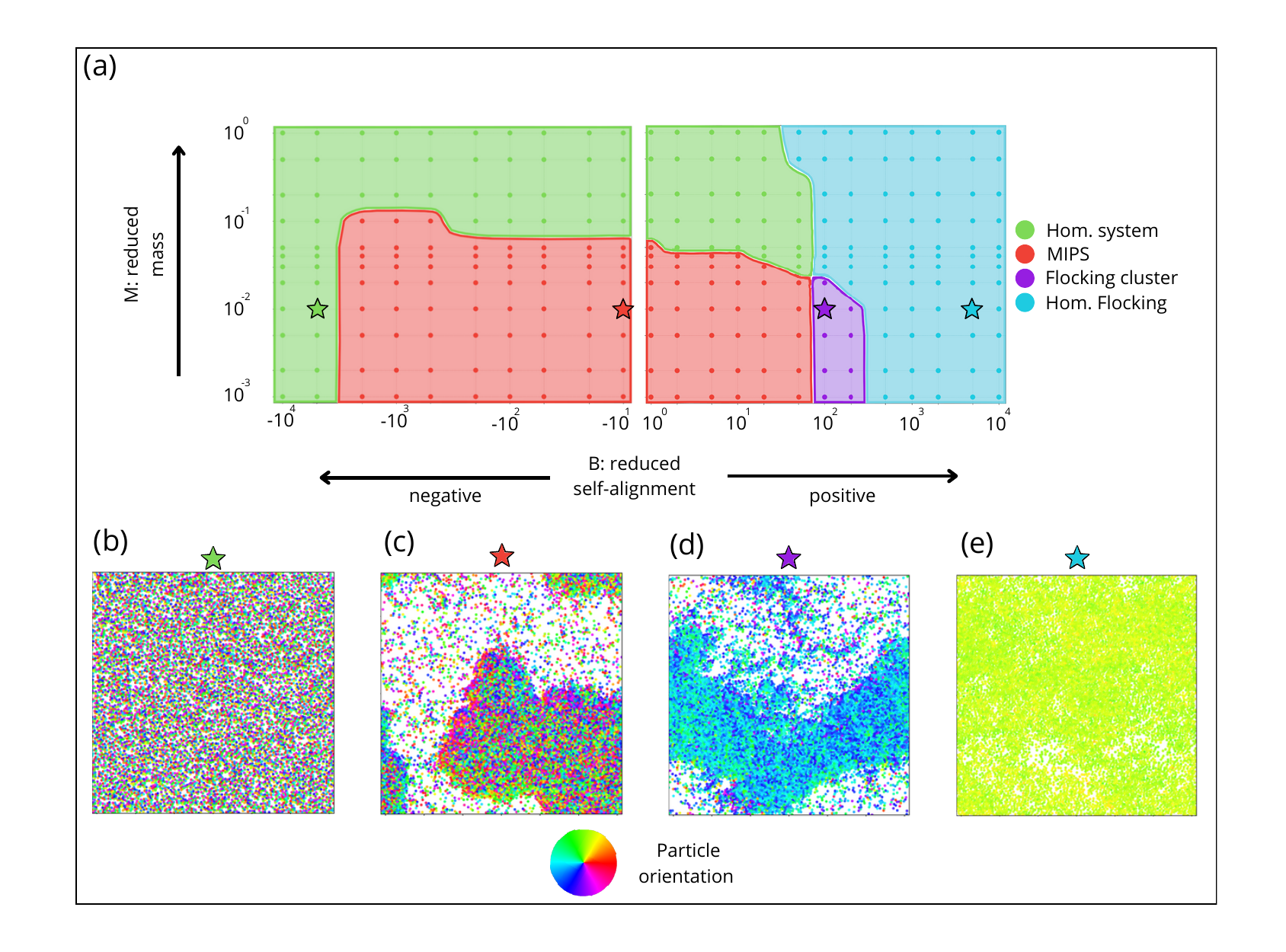}
\caption{
\textbf{Phase diagram.} (a) Phase diagram in the plane of reduced mass $M=D_r m/\gamma$ (inertial time) and reduced self-alignment time $B=\beta \sigma/(J D_r$). Positive values of $B$ correspond to self-alignment favoring orientations parallel to the velocities, while negative values of $B$ lead to anti-self-alignment inducing antiparallel orientations and velocities. Background colors are used to distinguish between different steady-state phases: a homogeneous phase (light green), Motility-induced phase separation (red), flocking motility-induced phase separation (violet), and a flocking homogeneous phase (light blue).   (b)-(e) Representative snapshots of the different phases corresponding to the stars in the phase diagram, where each particle is colored according to a color map based on its orientation. Simulations are realized at packing fraction $\Phi=0.5$ and number of particles $N=10^4$. The other dimensionless parameters of the simulations are $\text{Pe}=50$ and $D_r J/\gamma_r=10^{-3}$. 
    }
    \label{fig:fig2}
\end{figure*}

\begin{figure*}[t!]
    \centering
    \includegraphics[width=\textwidth]{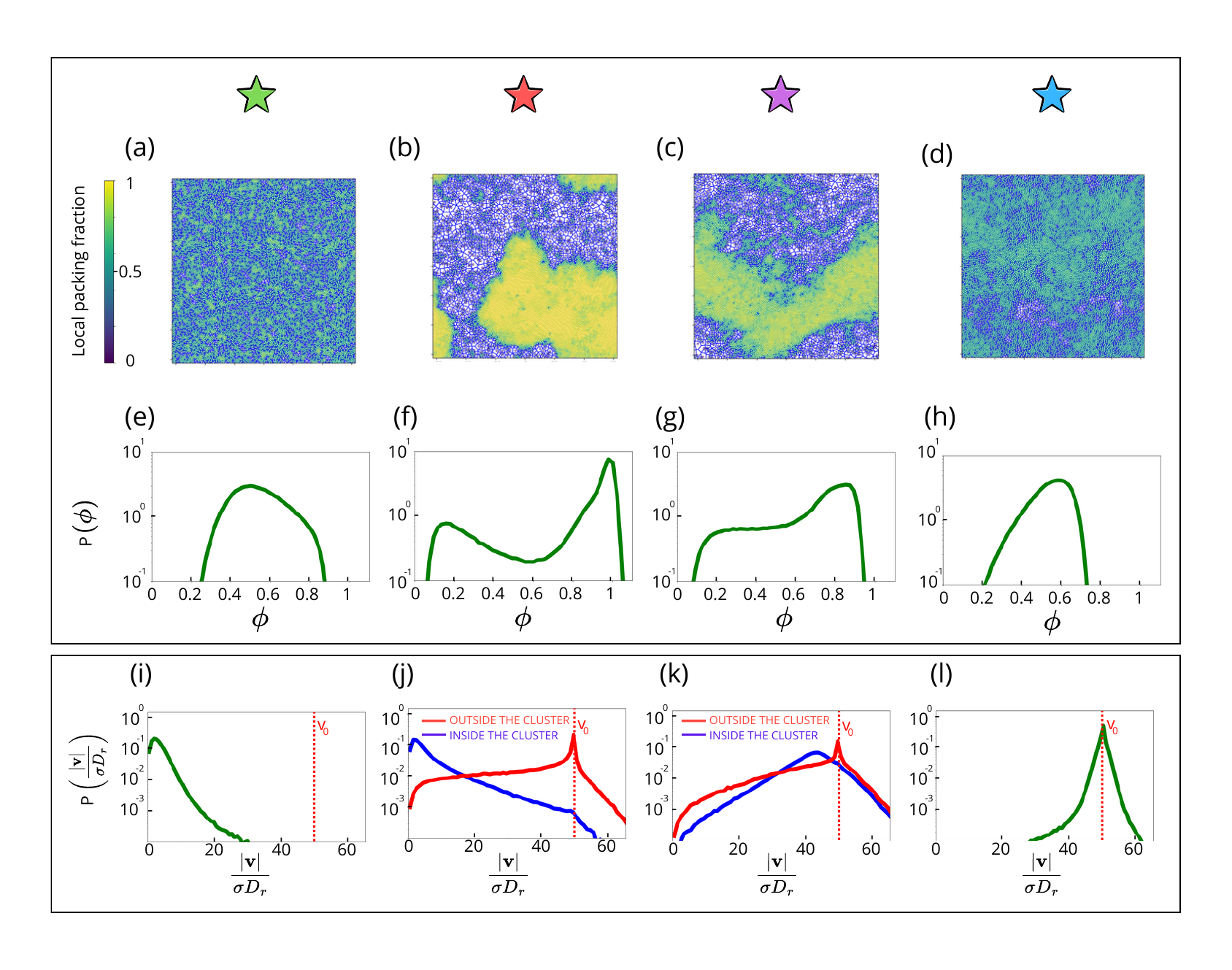}
        \caption{
            \textbf{Suppression of motility-induced phase separation.} 
            (a)-(d) Snapshots of simulations with color gradients representing the local packing fraction $\phi$, computed using the Voronoi tessellation. The stars coincide with those adopted in the phase diagram (Fig.~\ref{fig:fig2}~(a)) and are obtained at reduced mass $M=D_r m/\gamma_r=10^{-2}$ and reduced self-alignment strength $B=-5000, -10, 100, 5000$ for green, red, violet and light-blue colors, respectively.
            (e)-(h) Corresponding packing fraction distribution $P(\phi)$.
            (i)-(l) Speed probability distribution $P(|\mathbf{v}|)$ rescaled by $\sigma D_r$.
            In (i) and (l), green colors are used to denote observables calculated over the whole system, while, in (j)-(k), corresponding to phase-separated configurations, colors are used to distinguish between particles in the cluster (blue) and particles outside the cluster (red).
            Simulations are realized at packing fraction $\Phi=0.5$ and number of particles $N=10^4$. The other dimensionless parameters of the simulations are $\text{Pe}=50$ (represented by a vertical line in (i)-(l)) and $D_r J/\gamma_r=10^{-3}$.
    }
    \label{fig:fig3}
\end{figure*}

\subsection{Self- and anti-self alignment impact on a collision}

When two active Brownian particles without alignment or self-alignment mechanisms collide, their orientations remain independent. Particles cannot pass each other due to repulsive interactions and the collision impact induces two instantaneous velocities pointing away from each other. 
Being slowed down by the collision, the particle tends to restore the velocity before the collision due to the active force exerted toward the unaffected pre-orientational direction. This produces multiple interparticle impacts recently referred to as \textit{tapping collisions}~\cite{caprini2024emergent}. This scenario lasts until the self-propulsion vectors are reoriented for a time that is roughly determined by the persistence time $1/D_r$, after which the particle's orientation is altered.

The presence of the self-alignment mechanism between velocity and self-propulsion affects the collisions between two particles. Before the first impact, the two active particles move in the direction determined by their orientation, one against each other forming a small angle as in the first frame in Fig.~\ref{fig:fig1}~(c). 
In fact, velocity tends to follow the direction imparted by the active force (Eq.~\eqref{eq:eulero_pos}), and the orientation tends to align with the particle velocity as a direct effect of the self-alignment mechanism induced by the torque $\mathbf{T}^{sa}$ (Eq.~\eqref{eq:eulero_ang}).
After the impact, the particle velocity is sharply changed and they are roughly reversed. Subsequently, the velocity tends to align with the orientation (Eq.~\eqref{eq:eulero_pos}) and the orientation with the velocity through the self-alignment mechanism (Eq.~\eqref{eq:eulero_ang}).
However, the horizontal velocity component is suppressed due to ongoing collisions that occur when the particle velocities are again oriented towards each other (Fig.~\ref{fig:fig1}~(c), third frame). This mechanism allows only vertical velocity to which the orientation tends to slowly align until the two active particles have parallel velocities and orientations (See Video-1 in the Supplementary Material).
As a result, a configuration where the two active particles have parallel velocities and orientations is favored and, consequently, they move together with the same speed $|\mathbf{v}|\approx v_0$ as revealed by plotting the squared velocity as a function of time (dashed orange line in Fig.~\ref{fig:fig1}~(e)). Before the impact, the particles' velocities are equal to the free-particle velocity resulting from self-propulsion. After the collision, which causes them to reorient their motion parallel to each other, they regain this velocity. This final state, in which the particles move parallel to each other, is highlighted by the orange dashed line in Fig.~\ref{fig:fig1}~(f), where the relative angle between their orientational vectors approaches zero.
We note that the self-alignment mechanism between orientation and velocity is a single-particle effect. Therefore, in contrast with Vicsek models, indirect alignment between the velocities (and orientations) of different particles requires repulsive interactions, which prevents a direction of motion and favors a configuration where the two particles move together. Intuitively, such a mechanism could give rise to a global collective motion (flocking behavior) if the self-alignment strength $\beta$ is sufficiently large compared to the persistence time of the orientation and if a sufficiently large number of collisions occurs. 

The collision mechanism changes dramatically in the presence of anti-self-alignment, i.e.\ for negative values of $\beta$. Here, we have two competing mechanisms: the active force in Eq.~\eqref{eq:eulero_pos} that aligns velocity with the orientation, and the anti-self-alignment torque $\mathbf{T}^{sa}$ in Eq.~\eqref{eq:eulero_ang} that anti-aligns orientation with the velocity. 
When the anti-self-alignment torque dominates the dynamics, a particle immediately stops as the anti-self-alignment always turns the self-propulsion force opposite to the particle velocity. When the self-propulsion force dominates, the dynamics are more interesting.
Indeed, particles before a collision still tend to follow the respective orientations.  
However, after the impact when velocities are reversed, the anti-self-alignment torque pushes the particle's orientation to anti-align with the velocity (Fig.~\ref{fig:fig1}~(b)). As a consequence, this mechanism favors a configuration where each velocity forms an angle of $\pi$ radians with its orientation. In this case, self-propulsion also tends to suppress the velocity, as it acts in the opposite direction, favoring a configuration in which both particles eventually become stuck to each other with an almost vanishing velocity (See Video-2 in the supplementary material). This can be observed in the temporal evolution of $v^2$ in Fig.~\ref{fig:fig1}~(e) (solid blue line), where it is evident that the velocities of the particles approach zero after the collision. In the final configuration, the particles also orient against each other, resulting in a relative angle between their orientations that approaches $\pi$, as shown by the solid blue line in Fig.~\ref{fig:fig1}~(f).
This phenomenon intuitively suggests the emergence of a freezing effect, which is favored by interactions with the consequent suppression of motility-induced phase separation and clustering.

\subsection{Clustering suppression due to self-alignment and anti-self-alignment}

\begin{figure*}[t!]
    \centering
    \includegraphics[width=1\textwidth]{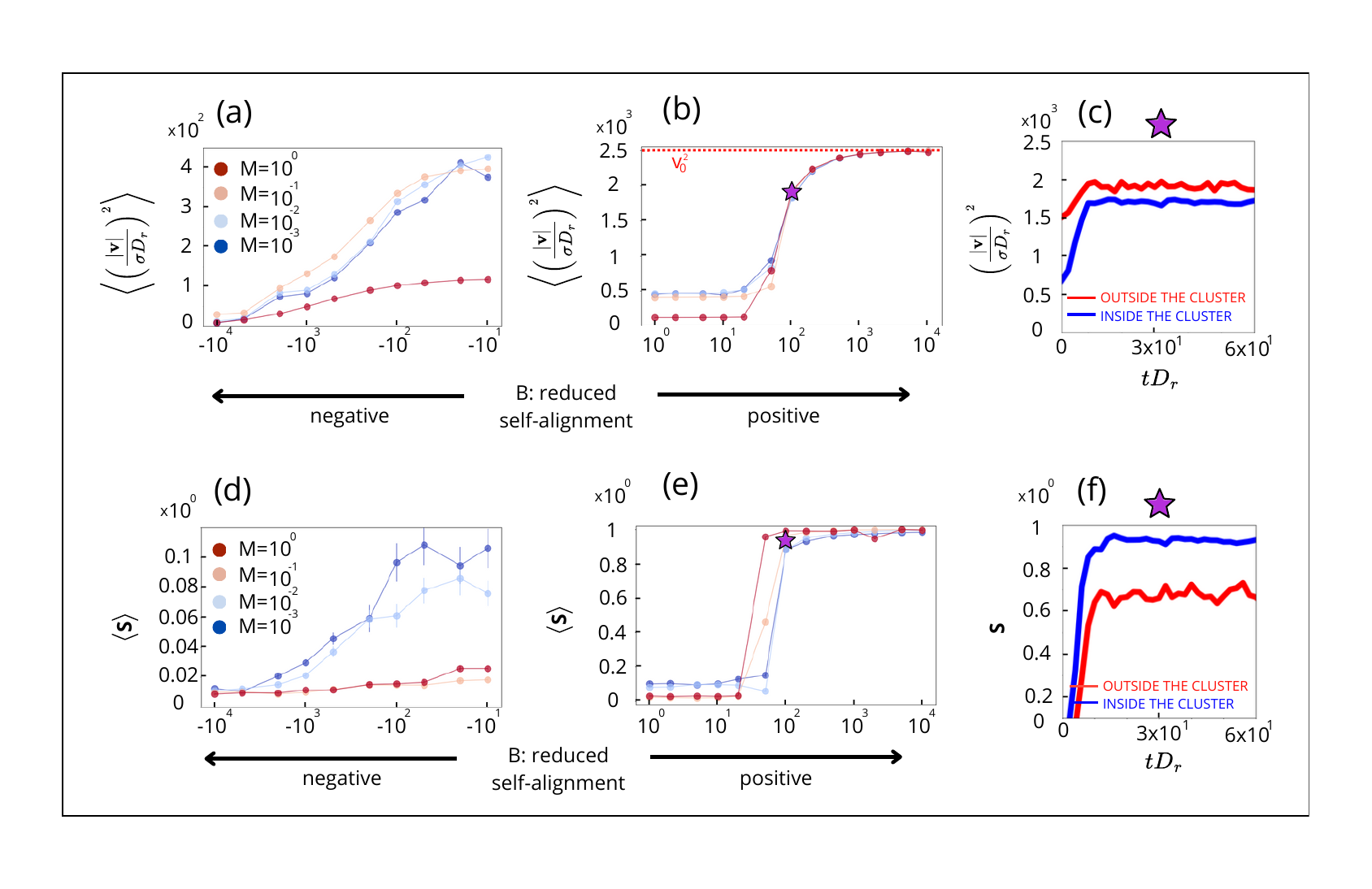}
       \caption{
% \justifying 
            \textbf{Flocking transition.}  
            (a)-(b) mean square velocity $\langle\mathbf{v}^2\rangle$ rescaled by $D_r^2 \sigma^2$ as a function of the self-alignment strength $B$ exploring negative (a) and positive values (b). Curves with different colors are obtained by considering different reduced inertial time $D_r m/\gamma$.
            (c) Time evolution of $\mathbf{v}^2$ (rescaled by $D_r^2 \sigma^2$) for $B= 100$ and reduced inertia $M=D_r m/\gamma=0.01$. This observable is calculated for particles in the high-density (blue) and low-density phases (red). 
            (d)-(e) average velocity-polarization $\langle S\rangle$ as a function of the self-alignment strength $B$ for negative (d) and positive (e) values. 
            $(f)$ time evolution of $S$ calculated inside (blue, high-density phase) and outside the cluster (red, low-density phase) at $B = 100$ and $M=D_r m/\gamma=0.01$. 
            Simulations are realized at packing fraction $\Phi=0.5$ and number of particles $N=10^4$.
            The other dimensionless parameters of the simulations are $\text{Pe}=50$ and $D_r J/\gamma_r=10^{-3}$.  
       }
    \label{fig:fig4}
\end{figure*}

The effect of self-alignment on collective phenomena is systematically explored in a phase diagram at packing fraction $\phi=0.5$ in the plane of the reduced self-alignment $B$ and reduced inertial time $m/\gamma D_r$ (Fig.~\ref{fig:fig2}~(a)).
When $B$ is positive, we explore self-alignment tending to align the orientation to the velocity.
By contrast, when $B$, through $\beta$, is negative, the particle motion is governed by anti-self-alignment, such that the particle orientation tries to orient anti-parallel to its velocity.

In the absence of a self-alignment mechanism ($\beta=0$), active Brownian particles with pure repulsive interactions display a non-equilibrium phase coexistence between a dense cluster and a dilute phase, termed motility-induced phase separation~\cite{cates2015motility}. 
This occurs in the absence of attractions due to the particle tendency of persistently moving in the same direction and remaining stuck against each other for small values of the reduced inertial time $m/\gamma D_r$.
This phase occurs for small reduced self-alignment strength $|B|$ (see for instance Fig.~\ref{fig:fig3}~(c)). Motility-induced phase separation is identified by monitoring the distribution of the local packing fraction $P(\phi)$. The phase coexistence previously described occurs when $P(\phi)$ displays a bimodal shape, with a first peak positioned at $\phi > \Phi$ and an additional peak at $\phi < \Phi$, identifying the high-density and the low-density phases. By contrast, in the homogeneous phase, $P(\phi)$ is characterized by a unimodal shape roughly peaked at $\phi \approx \Phi$. Consistently with previous results~\cite{mandal2019motility}, motility-induced phase separation is suppressed when the particle inertia is increased (Fig.~\ref{fig:fig2}~(a)).

Here, we discover that both self-alignment and anti-self-alignment promote a transition from a phase separated configuration (Fig.~\ref{fig:fig2}~(c)-(d)) to a homogeneous phase (see Fig.~\ref{fig:fig2}~(b) and Fig.~\ref{fig:fig2}~(e) for anti-self and self-alignment, respectively.
Phase-separated and homogeneous phases are qualitatively visualized by plotting coarse-grained snapshots colored according to the local packing fraction $\phi$ (Figs~\ref{fig:fig3}~(a)-(d)), while they are quantitatively identified by monitoring the distribution $P(\phi)$ which switches from a bimodal (Fig.~\ref{fig:fig3}~(f)-(g)) to a unimodal Fig.~\ref{fig:fig3}~(h)) shape.

%self-aligment: interpretation
The suppression of motility-induced phase separation detected, observed for self- ($B>0$) and anti-self-alignment ($B<0$), occurs for two different physical mechanisms. Let us start from the case $B>0$. The effective orientation and velocity alignment observed during a collision do not allow particles to remain stuck against each other. This is the opposite scenario compared to the mechanism responsible for motility-induced phase separation observed in non-aligning active particles. As a result, self-alignment alters the speed distribution, $P(|\mathbf{v}|)$.
For small self-alignment $B$ showing motility-induced phase separation, particles in the high-density phase are characterized by $P(|\mathbf{v}|)$ peaked at a small speed value (Fig.~\ref{fig:fig3}~(j)), while particles in the low-density phase display a $P(|\mathbf{v}|)$ roughly peaked at $v_0 \gg 0$ (Fig.~\ref{fig:fig3}~(j)-(k)). Indeed, particles in the cluster are stuck and exhibit limited mobility, while particles in the low-density phase move fast, nearly as free particles. As the self-alignment strength $B$ increases and motility-induced phase separation is suppressed, $P(|\mathbf{v}|)$ is narrowly distributed near $v_0 \gg 0$ (Fig.~\ref{fig:fig3}~(l)).
Despite the large packing fraction, particles move together as a single active object. Thus, active particles governed by self-alignment do not experience a density-dependent reduction in their effective velocity. Consequently, the typical theoretical explanation for clustering in repulsive active systems cannot be applied~\cite{cates2013active}.

The suppression of motility-induced phase separation occurs at a critical value of the reduced self-alignment strength $B$, which %weakly 
depends on the particle inertia. The larger the particle inertia, the smaller the self-alignment strength necessary to suppress the motility-induced phase separation (Fig.~\ref{fig:fig2}~(b)).
This result can be explained by noting that inertia slows down the velocity change by introducing a memory in the velocity evolution.

\subsection{Anti-self-alignment induces freezing}
%%% anti-self-alignment interpretation
The suppression of motility-induced phase separation induced by anti-self-alignment ($B<0$) can be explained by the collisional mechanism described in the previous section.
Indeed, anti-self-aligning particles push the particle orientation to become anti-parallel compared to the velocity vector. This mechanism effectively reduces the particle velocity, leading to a smaller effective persistence length for a single particle compared to $v_0/D_r$. 
As a consequence, the system behaves as if subject to an effective P\'eclet number smaller than $v_0/(\sigma D_r)$ until the system loses its capability of phase separating~\cite{Redner2013Structure, digregorio2018full, caprini2020hidden}. 

Our interpretation of the motility-induced phase separation suppression is based on a freezing effect directly caused by anti-self-alignment. To confirm this scenario, we monitor the typical speed distribution $P(|\mathbf{v}|)$ corresponding to large and small anti-self-alignment. In the former case, when motility-induced phase separation occurs, particles inside the cluster are characterized by a small speed, while particles in the dilute case move faster, roughly as free active particles with speed $v_0$ (Fig.~\ref{fig:fig3}~(j)), as occurs in athermal active particles. In the latter case, clustering does not occur and the speed distribution is peaked at values $|\mathbf{v}|\ll v_0$ (Fig.~\ref{fig:fig3}~(i)). 
Correspondingly, we observe that the mean square velocity $\langle |\mathbf{v}|^2\rangle$ is reduced, thereby confirming our physical interpretation of the suppression of motility-induced phase separation.

\subsection{Flocking transition and flocking motility induced phase separation}

%%% Introduction of the flocking order parameters
The speed distribution peaked at the single particle speed $v_0$ suggests that active particles align their orientation so that they can move in parallel, as shown by the collisional mechanism. When this process involves several particles, the system displays a flocking transition, as evidenced by monitoring the velocity polarization, $\langle S \rangle=\frac{1}{N}\langle|\sum_i \mathbf{v}_i/|\mathbf{v}_i||\rangle$ as an order parameter. In a disordered phase $\langle S \rangle\approx 0$, while in an ideal flocking phase where all the velocities are aligned, $\langle S\rangle\approx1$. We define the flocking when $\langle S \rangle$ exceeds the threshold value $\langle S \rangle>S_c \approx 0.8$.

%%% homogeneous flocking
Consistently with previous studies~\cite{paoluzzi2024flocking}, a homogeneous self-aligning liquid displays a transition from a disordered to a flocking phase when the self-alignment exceeds a critical value $\beta/\gamma_r\sim\beta_c/\gamma_r=D_r/v_0$, corresponding to the critical reduced self-alignment strength $B_c\sim\beta_c \sigma/(J D_r)$. When this happens, the time-trajectory of the velocity polarization $S$ approaches a steady-state plateau close to 1, while $\langle S\rangle$ as a function of $B$ displays the typical sigmoidal shape of a phase transition (Fig.~\ref{fig:fig4}~(e)). This transition occurs for several values of reduced inertia and is characterized by a small shift for large inertial time (Fig.~\ref{fig:fig2}~(a)).
In correspondence with the flocking phenomenon, the mean square velocity per particle $\langle |\mathbf{v}^2| \rangle$ increases as a function of the reduced self-alignment $B$ until the plateau $~v_0^2$ is reached. This value corresponds to the mean square velocity of a single active particle with speed $v_0$ and confirms that after aligning their orientations, self-aligning active particles flock (Fig.~\ref{fig:fig4}~(b)).

%%% flocking MIPS
Compared with previous results, we showcase that, before approaching the homogeneous-flocking configuration, the system displays a flocking motility-induced phase separation where the dense cluster flocks and the low-density phase do not. This is shown by monitoring the steady-state value for the order parameter $S$ inside and outside the cluster (Fig.~\ref{fig:fig4}~(f)). This study reveals a plateau value close to $1$ (flocking) for particles within the cluster and a lower plateau value for particles in the dilute phase.
This phase occurs for an optimal range of reduced self-alignment strength $B$, larger than the critical value of $B$ giving rise to a flocking homogeneous phase and smaller than another threshold value at which clustering is suppressed and the homogeneous flocking is recovered. As mentioned previously, interparticle interactions are fundamental to generating the alignment mechanisms between velocities and polarization of different particles. Indeed, high-density groups of interacting particles subject to multiple collisions align their velocity more easily than particles that rarely interact, as occurs in the low-density phase.
This explains why particles in the dense cluster have a larger tendency to align and display flocking, while particles in the dilute phase do not.
However, if this mechanism is too strong coherent motion is observed before any cluster nucleation and the system simply displays a homogeneous flocking phase.

%\subsection{scaling argument}
%scaling argument
The flocking transition observed for a homogeneous phase or a phase-separated configuration can be explained by considering a scaling argument. Active particles even in the absence of self-alignment are characterized by a persistence length, $v_0/D_r$, along which the particles proceed in the same direction roughly at speed $v_0$ for a time $1/D_r$ also termed persistence time. This typical length is quantified by the P\'eclet number which compares the persistence length with the particle diameter $\sigma$. In the presence of self-alignment, we can identify an additional typical length, which we term the self-alignment length, $\beta/\gamma_r$, which quantifies the distance needed by a particle to align the orientation to its velocity, to be compared with the particle diameter $\sigma$. This approximation holds in the limit of small rotational inertia adopted in this paper and can be obtained by multiplying the dimensionless parameter $B=\beta \sigma/(J D_r)$ with the reduced rotational inertia $D_r J/\gamma_r$ to obtain $\tilde{B}=\beta \sigma/\gamma_r$.

The transition from random to flocking behavior is roughly obtained when the product between $\tilde{B}$ exceeds the P\'eclet number,
$\tilde{B} \text{Pe} \gtrsim 1$. This condition implies that the typical length of the persistence length is larger than the typical length due to the self-alignment, such that
\begin{equation}
    \frac{v_0}{D_r} \gtrsim \frac{\gamma_r}{\beta} \,.
\end{equation}
Indeed, as explained by the collisional mechanism, flocking behavior can occur if the particles have the possibility to align their orientation to their velocity before the particles reorient. Consequently, the two particles should proceed in the same direction at least as long as the total duration of the self-alignment process. In other words, this collective effect occurs when the particle persistence length is larger than the typical length after which the polarization aligns with the velocity as an effect of the self-alignment mechanism.

%\vspace{1cm}
\section{Conclusions}

\noindent Here, we discover that both self-alignment and anti-self-alignment mechanisms suppress motility-induced phase separation (MIPS) in inertial active particles. The former promotes a flocking MIPS phase, where the cluster flocks in a random, low-density liquid before clustering is suppressed and a homogeneous flocking phase is recovered.
The latter induces a freezing phenomenon that reduces the effective particle motility until the particles are so slow that a cluster cannot form. Our study reveals that MIPS and flocking can effectively coexist.

These results are amenable for the experimental verification via active granular particles.
This result can be verified in experiments based on active granular particles, for instance, Hexbug particles, which are characterized by a degree of self-alignment. Alternatively, the strength of self-alignment can be tuned as desired by 3D-printing granular walkers with mass asymmetry in their body. We note that by using this strategy, it is even possible to design active granular particles governed by anti-self-alignment, by changing the position of the mass asymmetry compared to the particle center of mass.

Beyond the numerical exploration of the effect of self-alignment, future studies will aim to theoretically explain the MIPS suppression and the flocking transition due to self-alignment. Deriving a hydrodynamic description~\cite{Marconi_2021} and an effective field theory~\cite{te2023microscopic} from the microscopic dynamics could represent a theoretical step to analytically shed light on the flocking phase observed in liquids or lower density configurations.

\section{Acknowledgments} 
%LC acknowledges support from 
HL acknowledges support by the Deutsche Forschungsgemeinschaft (DFG) through the SPP 2265, under grant numbers LO 418/25.

\section*{AUTHOR DECLARATIONS}
\noindent
Conflict of Interest: The authors have no conflicts of interest to disclose.

\section*{DATA AVAILABILITY}
\noindent
The data that support the findings of this study are available from the corresponding author upon reasonable request.

\appendix
\begin{widetext}

\section{Details on the numerical simulations}
\label{appendix:A}
\noindent The dynamics described in the main text by \eqref{eq:eulero_pos} and \eqref{eq:eulero_ang}, regarding the time evolution of the particle's position, velocity, and orientation, are numerically implemented by using the Euler integration scheme with a time step of $\Delta t = 2 \times 10^{-5} \, \tau_p$. If we rescaled time by the persistence time of the particle's trajectory, $\tau_p$, and positions by the particle's diameter, $\sigma$, the integration scheme is as follows:
\begin{subequations}
\begin{align}
    \mathbf{r}'_i(t' + \Delta t') &= \mathbf{r}'_i(t') + \Delta t' \mathbf{v}'_i(t') \label{eq:x}\\[6pt]
    \mathbf{v}'_i(t' + \Delta t') &= \mathbf{v}'_i(t') - \frac{\gamma}{m D_r} \Delta t' \gamma \mathbf{v}'_i (t') + \Delta t' \frac{1}{\sigma m {D_r}^2} \mathbf{F'}_i (\mathbf{r}'_i(t')) + \frac{\gamma}{m D_r} \Delta t' v'_0 \mathbf{\hat{n}}_i \label{eq:v} \\[6pt]
    \theta'_i (t' + \Delta t') &= \theta_i' (t') + \Delta t' \omega'_i (t') \label{eq:phi} \\[6pt]
    \omega'_i (t' + \Delta t') &= \omega'_i (t') - \frac{\gamma_r}{J D_r}\Delta t' \omega'_i (t') + \frac{\beta \sigma}{J D_r} \Delta t' (\mathbf{\hat{n}}_i \times \mathbf{v'_i}) \cdot \hat{\mathbf{e}}_z + \frac{\gamma_r}{J D_r}\sqrt{2 \Delta t'} Y_i \label{eq:omega}
\end{align}
\end{subequations}
It is important to highlight that, in the equations presented, the use of the prime denotes the dimensionless quantities used in our simulations. In this notation, $\mathbf{\hat{n}}_i = (\cos(\theta_i), \sin(\theta_i))$, and $Y_i$ are Gaussian random numbers with zero mean and unit variance. Our dynamics are characterized by several dimensionless parameters, already introduced in the main text: the P\'eclet number, $\text{Pe} = v_0 / (D_r \sigma)$, which quantifies the activity strength and is fixed at $\text{Pe} = 50$ in this work; the reduced mass, $M = D_r m / \gamma$, which can be extracted as the inverse of the factor that multiplies the second and the last terms on the right-hand side of equation~\eqref{eq:v}, and which is varied between $10^{-4}$ and 1. By correctly treating the third term on the right-hand side of the same equation, we can extract another dimensionless parameter given by $\sqrt{\epsilon/m}/(D_r \sigma)$. The reduced moment of inertia, $I = D_r J / \gamma_r$, is fixed at $I = 0.001$ for all simulations and can again be extracted as the inverse of the factor that multiplies the second and the last terms on the right-hand side of equation~\eqref{eq:omega}. Finally, the reduced (anti-) self-alignment term, $B = \beta \sigma / (J D_r)$, is varied from $-10^4$ to $10^4$ and can be identified as the factor that multiplies the third term on the right-hand side of equation~\eqref{eq:omega}.
The final simulation time depends on the simulation parameters but is always around 2x$10^{2} \, \tau_p$. The simulations were performed in a system composed of $N = 10^4$ particles in a square box of size $L = 125$, ensuring that the packing fraction is equal to $\Phi = N \pi (\sigma/2)^{2} / L^{2} = 0.5$.
 
\end{widetext}

%\end{widetext}

%\bibliographystyle{mdpi}
%\bibliographystyle{rsc} %the RSC's .bst file
\bibliographystyle{rsc} %the RSC's .bst file

\bibliography{bib.bib}

%%%%%%%%%%%%%%%%%%%%%%%%%%%%%%%%%%%%%%%%%%
%% optional
%\sampleavailability{Samples of the compounds ...... are available from the authors.}

\end{document}